\documentclass[12pt]{article}
\usepackage{amsmath,amsfonts, amssymb, braket, bbold}
\usepackage{tikz}
\usetikzlibrary{trees,er,snakes,shapes,mindmap}
\usepackage{titlesec}
\titleformat{\section}
 {\normalfont\fontfamily{put}\fontsize{12pt}{16pt}\bfseries\color{black}}
{\thesection}{1em}{}
\titleformat{\subsection}
 {\normalfont\fontfamily{put}\fontsize{12pt}{16pt}\bfseries\color{black}}
{\thesubsection}{1em}{}
\def \beq  {\begin{equation}}
\def \eeq  {\end{equation}}
\def \beqar {\begin{eqnarray}}
\def \eeqar {\end{eqnarray}}
\allowdisplaybreaks
\def\sqr#1#2{{\vcenter{\vbox{\hrule height.#2pt
\hbox{\vrule width.#2pt height#1pt \kern#1pt
\vrule width.#2pt}\hrule height.#2pt}}}}

\def\la {{\langle}}
\def\ra {{\rangle}}

\def\vk {{\vec k}}

\def\Tr {{\rm Tr}}
\def \tr {{\rm tr}}

\def\bD {\bar{D}}

\def\vk {\vec{k}}

\def\del {\partial}

\def\a {\alpha}
\def\b {\beta}
\def\e {\epsilon}

\def\A {{\cal A}}

\def\F {{\cal F}}

\def\I {{\cal I}}

\def\half{\textstyle{1\over 2}}
\def\quarter {\textstyle{1\over 4}}

\def\la{{\langle}}
\def\ra{{\rangle}}

\mathchardef\mhyphen="2D
\begin{document}
\fontfamily{bch}\fontsize{12pt}{16pt}\selectfont
\def \CMP {{Commun. Math. Phys.}}
\def \PRL {{Phys. Rev. Lett.}}
\def \PL {{Phys. Lett.}}
\def \NPBProc {{Nucl. Phys. B (Proc. Suppl.)}}
\def \NP {{Nucl. Phys.}}
\def \RMP {{Rev. Mod. Phys.}}
\def \JGP {{J. Geom. Phys.}}
\def \CQG {{Class. Quant. Grav.}}
\def \MPL {{Mod. Phys. Lett.}}
\def \IJMP {{ Int. J. Mod. Phys.}}
\def \JHEP {{JHEP}}
\def \PR {{Phys. Rev.}}
\def \JMP {{J. Math. Phys.}}
\def \GRG{{Gen. Rel. Grav.}}
\begin{titlepage}
\null\vspace{-62pt} \pagestyle{empty}
\begin{center}
\vspace{1.3truein} {\large\bfseries
Transport coefficients for higher dimensional quantum Hall effect}
\vskip .1in
{\Large\bfseries ~}\\
\vskip .2in
{\sc Dimitra Karabali$^{a,c}$, V.P. Nair$^{b, c}$}\\
\vskip .2in
{\sl $^a$Physics and Astronomy Department,
Lehman College, CUNY\\
Bronx, NY 10468}\\
\vskip.1in
{\sl $^b$Physics Department,
City College of New York, CUNY\\
New York, NY 10031}\\
\vskip.1in
{\sl $^c$The Graduate Center, CUNY\\
New York, NY 10016}\\
 \vskip .1in
\begin{tabular}{r l}
{\sl E-mail}:&\!\!\!{\fontfamily{cmtt}\fontsize{11pt}{15pt}\selectfont 
dimitra.karabali@lehman.cuny.edu}\\
&\!\!\!{\fontfamily{cmtt}\fontsize{11pt}{15pt}\selectfont vpnair@ccny.cuny.edu}\\
\end{tabular}
\vskip .5in

\centerline{\large\bf Abstract}
\end{center}
An effective action for the bulk dynamics of quantum Hall effect in arbitrary even spatial dimensions was obtained some time ago in terms of a Chern-Simons term associated with the Dolbeault index theorem.
Here we explore further properties of this action, showing how electronic band structures can be incorporated, obtaining Hall
currents and conductivity (for arbitrary dimensions)
in terms of integrals of Chern classes 
for the bands. We also derive the expression for Hall viscosity
from the effective action. Explicit formulae for the Hall viscosity are given
for 2+1 and 4+1dimensions.

\end{titlepage}
\fontfamily{bch}\fontsize{12pt}{17pt}\selectfont
\pagestyle{plain} \setcounter{page}{2}
\section{Introduction}
Quantum Hall effect has been intensively investigated for several decades by now from both theoretical and experimental points of view \cite{QHE-general}.
An interesting variant has been its generalization to higher dimensions
\cite{ZH}-\cite{everyone}.
Even though seemingly this is only of mathematical interest, 
it is intriguing that this may in fact be experimentally realizable
using the idea of synthetic dimensions \cite{{4DQHE}, {4DQHE1}}.
Shortly after the initial work on higher dimensional QHE \cite{ZH}, it was realized that complex manifolds present a class of spaces for which one can explicitly solve the Landau problem, construct states, analyze edge excitations, etc. and therefore uniformly extend QHE to all even spatial dimensions
\cite{KN1}-\cite{KN4}.
Since holomorphicity is the key feature for states in the lowest Landau level,
the Dolbeault index theorem provides a convenient mathematical technique
for analyzing the phenomenon in arbitrary dimensions \cite{EGH}.
Some time ago, we used this connection to construct the {\it topological} effective bulk action for a quantum Hall system {\it of integer filling fraction} in arbitrary even dimensions, including both gauge and gravitational fluctuations,
in terms of a Chern-Simons action associated to the Dolbeault index density
\cite{KN}. We were able to construct such effective actions for arbitrary integer filling fraction and Abelian and non-Abelian gauge fields, the latter being a novel possibility in higher dimensions.
We may note here that effective actions in 2+1 dimensions, including gravitational contributions,
have been constructed by many authors, see, for example,
\cite{WZ}-\cite{AG2}.

In this paper we will investigate the general effective action obtained in \cite{KN} further focusing on the derivation of response functions and relevant transport coefficients in higher dimensions. For simplicity we will focus on the case of Abelian gauge fields but arbitrary even spatial dimensions. The response of the system to gauge and gravitational fluctuations is characterized by the electromagnetic current and the energy-momentum tensor which can be straightforwardly derived from the effective action.  

The transport coefficient related to the electromagnetic current is the Hall conductivity expressed in terms of the filling fraction. In the construction of the effective action from the Dolbeault index, the filling fraction was one of the input ingredients. However, in
deriving the effective action we essentially considered an appropriate gauge-covariant Laplacian as the Hamiltonian for the Landau problem, which means that we formulated the action in terms of Landau levels for fermions in free space. More realistically though
the fermions correspond to extended states in an energy band in the material. The Hall current and the filling fraction relevant to the quantum Hall state should thus be expressed in terms of the integrals of the Chern classes for the Berry curvature of the bands, as was done long ago for the two-dimensional case \cite{TKNN}.
This is particularly important if we seek experimental realizations for the higher dimensional cases. This is the first problem we address
 in this paper. We show how the band structure can be incorporated 
 in the effective action.
 The electromagnetic currents we obtain by this method agree with 
results obtained for 4d and 6d QHE using Hamiltonian perturbation theory for wave packets \cite{currents}.
Further we obtain explicit expressions for the currents in arbitrary even dimensions including contributions due to the spatial curvature.

Another transport coefficient of interest is the Hall viscosity.
This is
obtained from the response to perturbations of the
metric, from the two-point function for the energy-momentum tensor
\cite{{visc},{AG1}}.
We show how the Hall viscosity can be derived from the effective action, giving explicit formulae for 2+1 and 4+1 dimensions.
While the Hall viscosity has been derived in 2+1 dimensions by 
explicit calculation of the responses, our derivation places it within a uniform procedure easily applicable in any number of dimensions.
\section{Review of the derivation of the effective action using an index theorem }
In this section we give a brief resume of the bulk effective action and how it is
obtained from the index theorem \cite{KN}.
Although we may need to consider general
perturbations of the metric later, to begin with, 
the spatial manifold of interest  for us is
a complex K\"ahler manifold, such as $\mathbb{CP}^k$.
The single particle Hamiltonian is of the form $D_i \bD_{\bar i}$,
where $D_i$ and $\bD_{\bar i}$ are the holomoprhic and antiholomorphic covariant derivatives which include background gauge and gravitational fields.
 Thus the
wave functions for the lowest Landau level obey the holomorphicity
condition
\beq
\bD_{\bar i} \Psi = 0
\label{band12}
\eeq
The number of normalizable solutions to this equation is given by the index theorem for the twisted Dolbeault complex
as
\beq
{\rm Index}( \bD ) = \int_K {\rm td}(T_cK) \wedge {\rm ch} (V)
\label{band13}
\eeq
where ${\rm td}(T_cK)$ is the Todd class on the complex tangent space of $K$
and ${\rm ch}(V)$ is the Chern character of the relevant
vector bundle \cite{EGH}. 

An explanation of the various terms and terminology in (\ref{band13}) might be
useful before we proceed.
Generally the spin
connections and curvatures take values in the Lie algebra of the holonomy group, which is $SO(2k)$ for a real manifold in $2k$ dimensions.
For a complex manifold, coordinate transformations which preserve the complex structure are holomorphic transformations. This restricts the
holonomy group to $U(k) \subset SO(2k)$.
Correspondingly, the frame fields can be taken to be holomorphic and antiholomorphic one-forms which are combinations of the real ones
given by the complex structure. The tangent space
also has similar combinations which give $T_cK$.
The Todd class is given in terms of the curvature two-form for
$T_c K$.
It has the expansion \cite{EGH}
\beq
{\rm td} = 1 + {1\over 2} \, c_1 +{1\over 12} ( c_1^2 + c_2) + {1\over 24} c_1\, c_2
+ {1\over 720} ( - c_4 + c_1\,c_3 + 3 \, c_2^2 + 4\, c_1^2 \,c_2 - c_1^4) + \cdots
\label{band14}
\eeq
where $c_i$ are the Chern classes. For any vector bundle with curvature ${\cal F}$, the Chern classes are defined by\footnote{We start with connections and curvatures in an antihermitian basis since they are natural
allowing us to write $F = d A + A A$, etc. This leads to some factors
of $i$ in various expressions at this stage. Later we will move to a hermitian basis.}
\beq
\det \left( 1 + {i \, {\cal F} \over 2 \pi} \,t\right) = \sum_i c_i \, t^i
\label{band15}
\eeq
Various terms in the expansion (\ref{band14}) can thus be expressed as
powers of the curvature two-form.
The Todd class may also be represented, via the splitting principle,
in terms of a generating function as
\beq
{\rm td} = \prod_i {x_i \over 1- e^{-x_i}}
\label{band16}
\eeq
where $x_i$ represent the ``eigenvalues" of the curvature in a suitable canonical form
(diagonal or the canonical antisymmetric form for real antisymmetric $i\,{\cal F} $). The expansion in (\ref{band14}) is obtained by using this generating function and rewriting it using traces of powers of curvatures.

In the case of the index as in (\ref{band13}),
$\F$ is to be taken as the curvature two-form $R$ for
$T_cK$.
The first few Chern classes for the complex tangent space can then
be explicitly written, using (\ref{band15}), as
\beqar
c_1 (T_c K)& = & \Tr ~{iR \over 2\pi} \nonumber \\
c_2 (T_c K)& = & {1 \over 2} \Biggl[ \Bigl(\Tr {iR \over 2\pi}\Bigr)^2 - \Tr \Bigl({iR \over 2\pi}\Bigr)^2 \Biggr] \nonumber \\
c_3 (T_c K) &=& { 1 \over 3!} \Biggl[ \Bigl(\Tr {iR \over 2\pi}\Bigr)^3 - 3\, \Tr {iR \over 2\pi} \,\Tr \Bigl({iR \over 2\pi}\Bigr)^2 + 2 \,\Tr \Bigl({iR \over 2\pi}\Bigr)^3 \Biggr] \label{band16a}\\
c_4 (T_cK) &=& { 1 \over 4!} \Biggl[ \Bigl(\Tr {iR \over 2\pi}\Bigr)^4 - 6 \Bigl(\Tr {iR \over 2\pi}\Bigr)^2 \,\Tr \Bigl({iR \over 2\pi}\Bigr)^2 + 8\, \Tr {iR \over 2\pi}\, \,\Tr \Bigl({iR \over 2\pi}\Bigr)^3\nonumber\\
&&\hskip .4in  +3\, \Tr \Bigl({iR \over 2\pi}\Bigr)^2 
\,\Tr \Bigl({iR \over 2\pi}\Bigr)^2 -6\,\Tr \Bigl({iR \over 2\pi}\Bigr)^4 \Biggr] \nonumber
\eeqar
The curvatures $R$ take values in the Lie algebra of $U(k)$, which is the holonomy group for a complex manifold of real dimension $2k$.
The traces in the above formula are thus over the $U(k)$ Lie algebra.

Turning to the other set of terms in the index formula (\ref{band13}),
we may first note that the terminology of
the Dolbeault complex being twisted, as often used in mathematics literature,
is equivalent to saying that we have background gauge fields. 
In other words,
the vector bundle $V$ relevant for us is defined by the internal gauge symmetry
structure of the fermions fields. The fermion
wave functions are sections of this bundle.
The Chern character is given by
\beq
{\rm ch}(V) = \Tr \left( e^{i {\cal F} /2 \pi} \right) = {\rm dim}\,V + \Tr ~{{i {\cal F}} \over {2\pi}} + { 1 \over 2!} \Tr~ {{i{\cal F} \wedge i {\cal F}} \over {(2\pi)^2}} + \cdots
\label{band17}
\eeq 
where ${\rm dim} V$ is the dimension of the bundle $V$.
Here ${\cal F}$ is the gauge field strength $F$.
If spin is included,  ${\cal F}$ will also
include the curvature of the spin bundle.

By definition, the Dolbeault index gives the degeneracy of the lowest Landau level. If we consider a fully filled Landau level with filling fraction $\nu=1$ and assign a unit charge to each fermion, the index will also give the total charge. The charge density may therefore be identified with the index density allowing us to construct an effective action.
Thus in terms of an effective action $S_{\rm eff}$, we can then write
\beq
{\delta S_{\rm eff} \over \delta A_0} = J_0 =  {\rm Index ~density}
\label{band18}
\eeq
This shows that the leading term of the effective action may be taken as
a Chern-Simons term $CS(A)$ whose variational derivative with  respect to
$A_0$ will give the index density. In other words, we can
``integrate up" the relation (\ref{band18}), {\it and appropriately covariantize} to obtain {\it the topological part of the action}
$S_{\rm eff}$. (There can be subleading terms arising from
dipole and higher multipole terms in $J_0$ which integrate to zero
in the total charge.
They can contribute terms involving derivatives of the fields
in the effective action; they are also nontopological in nature.
The Chern-Simons form associated to the index density
 is the leading term in the sense of a derivative expansion,
see \cite{KN} for more details on this point.)
This procedure, as described so far,
does not determine the purely gravitational terms in
$S_{\rm eff}$. 
However, even though we are interested in the bulk action, one could envisage the situation of a droplet with edge modes; these will generate
a gravitational anomaly. The purely gravitational terms in $S_{\rm eff}$ will be determined by the gravitational anomaly {\it via the descent method used for anomalies} \cite{anom}.
This was the procedure we used in \cite{KN} to obtain the effective bulk
action
for higher dimensional quantum Hall effect.

In the case of a fully filled lowest Landau level we derived a compact form for the topological effective action for a general complex manifold of even spatial dimension $2k$, given by
\beq
S_{\rm eff} =
 \int \Bigl[ {\rm td}(T_c K) \wedge \sum_p  (CS)_{2 p+1} ( A)\Bigr]_{2 k+1}
+ 2 \pi\int \Omega^{\rm grav}_{2k+1}
\label{band18a}
\eeq
Here $(CS)_{2 p +1}(A)$ is the Chern-Simons term associated with
just the gauge part and is defined by
\beq
{1\over 2 \pi} d (CS)_{2 p +1} = {1 \over (p+1)!} \Tr \left(
{ i F \over 2 \pi}\right)^{p +1}
\label{band18b}
\eeq
One should expand the terms in the square brackets in
(\ref{band18a}) in powers of
curvatures and $F$ and pick out the term corresponding to
the $(2k+1)$-form. This is indicated by the subscript $2k+1$ for
the square brackets. The purely gravitational term 
$\Omega^{\rm grav}_{2k+1}$  in (\ref{band18a}) is defined by
\beq
\left[ {\rm td}(T_cK) \right]_{2 k +2}
=  d\, \Omega_{2 k+1}^{\rm grav} 
\label{band18c}
\eeq
Notice that here we start with the $(2k+2)$-form and define the
appropriate Chern-Simons form. The justification for this is via 
the well-known descent equations for anomalies, see \cite{KN, anom}.

These results can be further extended to include higher Landau levels for some
special cases. For many manifolds, such as $\mathbb{CP}^k$,
the degeneracy for spinless fermions
in the $s$-th Landau level is identical to the degeneracy for
fermions of spin $s$ in the lowest Landau level. One can then use
the index theorem as before to construct the effective action.
The procedure is exactly as outlined above, except that $\F$ in the Chern character now includes the curvature for the spin bundle as well.
Explicitly, what this means is that
\beq
\F = F + {\cal{R}}_s = F + s\, R^0 \mathbb{1} + R^aT_a
\label{band18d}
\eeq
where $R^0$ and $R^a$ denote the curvatures for the
$U(1)$ and $SU(k)$ factors in $U(k) \subset SO(2k )$.
$T_a$ are the generators of $SU(k)$ in the appropriate
representation of the appropriate spin.
The general effective action has the form
\beq
S^{(s)}_{\rm eff} =
 \int \Bigl[ {\rm td}(T_c K) \wedge \sum_p  (CS)_{2 p+1} (\omega_s + A)\Bigr]_{2 k+1}
+ 2 \pi\int \Omega^{\rm grav}_{2 k+1}
\label{band18e}
\eeq
where $\omega_s$ is the spin connection for 
$s R^0 \mathbb{1} + R^aT_a$.

Specifically, the effective actions for $2+1$, $4+1$ and $6+1$ dimensions were
worked out in detail. In $2+1$ dimensions, the result for the $s$-th Landau level is
\beqar
S_{2+1}^{( s)} &=& {i^2 \over {4\pi}}\left[  \int A \Bigl(dA +2\Bigl(s+{1 \over 2}\Bigr)d\omega^0 \Bigr) +   \Bigl( \Bigl(s+{1 \over 2}\Bigr)^2 - {1\over 12} \Bigr)
\int  \omega^0 \, d \omega^0 \right] \nonumber \\
&=& {i^2 \over {4\pi}} \int \Biggl\{\Bigl[A +(s+{1 \over 2})\,\omega^0\Bigr]d\Bigl[A +(s+{1 \over 2})\,\omega^0\Bigr] - {1 \over 12} \omega^0 \,d\omega^0 \Biggr\}
\label{band19}
\eeqar

In $4+1$ dimensions, we find
\beqar
S_{4+1}^{(s)} &=&{i^3 (2j+1)\over {(2\pi)^2}} \int \Biggl\{ { 1 \over 3!} \Bigl(A+(s+1)\omega^0\Bigr) \Bigl[d\Bigl(A+(s+1)\omega^0\Bigr)\Bigr]^2 \nonumber \\
&&-{ 1 \over 12} \Bigl(A+(s+1)\omega^0\Bigr) \Biggl[  (d\omega^0)^2 -\Bigl[ (4j(j+1) - {1\over 2} \Bigr] {1 \over 2} ( {R^a} \wedge { R^a}) \Biggr] \Biggr\}
\label{band20}
\eeqar
where $j = s/2$.
For a complex manifold such as $\mathbb{CP}^2$ in four dimensions,
the holonomy group (which is where the gravitational curvatures and connections take values)
 is $U(2) \subset SO(4)$.
 In (\ref{band20}), $\omega^0$ denotes the $U(1)$ spin connection,
 with $U(2) \sim SU(2) \times U(1)$, and ${R^a}$ is the curvature
 for $SU(2)$. 

In $6+1$ dimensions, the action for the lowest Landau level
($s = 0$) was obtained as
\beqar
S_{6+1}^{(0)} &=& { 1 \over (2\pi)^3} \int \Biggl\{ { 1 \over 4!} \left(A + {3 \over 2} \omega^0\right) \left[ d\left( A + {3 \over 2} \omega^0\right)\right]^3\nonumber\\
&&\hskip .7in - {1 \over 16} \left(A + {3 \over 2} \omega^0\right)d\left(A + {3 \over 2} \omega^0\right) \left[ (d\omega^0)^2 + {1 \over 3} \Tr ({\tilde R} \wedge {\tilde R}) \right] \nonumber \\
&&\hskip .7in +{ 1 \over 1920} \omega^0d\omega^0 \left[ 17 (d\omega^0)^2 + 14\, \Tr ({\tilde R} \wedge {\tilde R}) \right] + {1 \over 720} \omega^0 \Tr ({\tilde R} \wedge {\tilde R}\wedge {\tilde R}) \Biggr\} 
\nonumber\\
&&+ {1 \over 120} \int (CS)_7 ({\tilde \omega})
\label{band21}
\eeqar
Here we are only displaying the result for the lowest Landau level
for simplicity. Similar to the $4+1$-case, we have a $U(1)$ subgroup
of the holonomy group with spin connection $\omega^0$ and
an $SU(3)$ subgroup with spin connection ${\tilde \omega}$.
The curvature ${\tilde R}$ corresponds to
${\tilde \omega}$; it is in the fundamental representation of
$SU(3)$, as given in (\ref{9A}), (\ref{9a}) in Appendix A. Also
$(CS)_7 ({\tilde\omega})$ in (\ref{band21})  is the standard Chern-Simons $7$-form for this spin connection.

\section{Hall currents and characteristic classes for electronic bands}
In this section, we will describe how the effective actions derived
in \cite{KN} (relevant for electrons in free space) 
should be modified to incorporate
the geometrical properties of the electronic bands.
We start by reiterating that the expressions for the effective actions (\ref{band18a}), (\ref{band18e}) are valid for both Abelian and non-Abelian gauge fluctuations. By construction, the variation of the effective action
is of the form
\beqar
\delta S_{\rm eff}&=& \int \delta A\, \I\nonumber\\
\I&=&  \sum_{l=0}^k {1\over (k-l)!} \left( {F \over 2 \pi}\right)^{k-l} \wedge
\Bigl[{\rm td}(T_cK)\Bigr]_{2l}
\label{SLLL}
\eeqar
Here $\I$ is a differential $2k$-form. Formally it is identical to the index density, but here we are interpreting it as a
form in the $(2k+1)$-dimensional spacetime. Thus it defines a dual
vector, which is the Hall current.
In the following, for simplicity, we shall only consider Abelian gauge fields and consider the electromagnetic Hall current. 
We will further focus on the lowest Landau level, $s = 0$.

The expressions for the Hall current obtained as in
(\ref{SLLL}) correspond to fermions in free space. 
To work out the required modification taking account of the fact that the fermions are from an energy band appropriate to the material,
we consider the motion of a particle, viewed as a wave packet with position $x^i$ and momentum
$k^i$. The dynamics of such a wave packet is described by the action
\cite{semi}
\beq
S = \int k_i {\dot x^i} - E(k)  + e \phi (x) - A_i {\dot x}^i
+ \A_i {\dot k}^i
\label{band1}
\eeq
The wave packet is to be viewed as describing a  single-particle extended state within an energy band, with energies given by
$E(k)$. Here $\phi$ is the electrostatic potential,
$A_i$ is the magnetic vector potential  and $\A_i$ is the Berry connection defined by
\beq
\A_i (k) = \int [dx]\, \Psi_k^\dagger (x) {\del \over \del k^i} \Psi_k(x)
\label{band2}
\eeq
We are considering a $2k$-dimensional spatial manifold and
$[dx]$ in (\ref{band2}) gives
the appropriate volume element.
The canonical symplectic structure associated with the action (\ref{band1}) is easily read off as
\beq
\omega_{\rm symp} = dk_i\, dx^i - {1\over 2} F_{ij} dx^i \, dx^j
+ {1\over 2} \Omega_{ij} dk^i \, dk^j
\label{band3}
\eeq
where the gauge and Berry curvatures are given by
\beqar
F_{ij} &=& {\del \over \del x^i} A_j - {\del \over \del x^j} A_i \nonumber\\
\Omega_{ij} &=& {\del \over \del k^i} \A_j - {\del \over \del k^j} \A_i
\label{band4}
\eeqar
The canonical structure (\ref{band3}) shows that the commutators
$[x^i, x^j]$ and $[k_i, k_j]$ will be nonzero since there are 
$dk^i\, dk^j$ and $dx^i\, dx^j$ terms in $\omega_{\rm symp}$, in addition to
the standard term $dk_i\, dx^i$.
Our aim now will be to choose variables so as to
eliminate mixing between the two sectors.
Towards this we write
\beq
k_i = \lambda_{ij} p^j + \sigma_{ij} x^j
\label{band5}
\eeq
This introduces $p_i$ which will take the place of
$k_i$.
In the following, we will consider $F_{ij}$ and $\Omega_{ij}$ to be approximately constant; this is appropriate to the lowest order in what may be considered as a gradient expansion.
With the substitution (\ref{band5}), $\omega_{\rm symp}$ becomes
\beqar
\omega_{\rm symp} &=&  \left[ \lambda  -\sigma^T \Omega \lambda 
\right]_{ij} \,
dp^i \, dx^j \nonumber\\
&&+
\left[ - \sigma - {1\over 2} F + {1\over 2} \sigma^T \Omega \sigma
\right]_{ij} dx^i \, dx^j
+ {1\over 2} \left[ \lambda^T \Omega \lambda \right]_{ij} dp^i \, dp^j
\label{band6}
\eeqar
The term mixing $x^i$ and $p^i$ can be eliminated by imposing
\beq
\lambda  -\sigma^T \Omega \lambda = 0
\label{band7}
\eeq
The solution to this equation is given by
$\sigma = - \Omega^{-1}$.
We can then simplify $\omega$ as
\beq
\omega_{\rm symp} = {1\over 2} \left( \Omega^{-1} - F\right)_{ij} dx^i \, dx^j
+ {1\over 2}  \left( \lambda^T \Omega \lambda \right)_{ij} dp^i \, dp^j
\label{band8}
\eeq
We have separated the $x$- and $p$-dependent terms, but
$\lambda$ is not determined by our considerations so far.
We will make the simplest choice that it is the identity matrix; this will suffice for our purpose. In this case, we have
\beq
\omega_{\rm symp} = {1\over 2} \left( \Omega^{-1} - F\right)_{ij} dx^i \, dx^j
+ {1\over 2} \Omega_{ij} dp^i \, dp^j
\label{band9}
\eeq
Recall that the lowest Landau level dynamics  for fermions in free space coupled to a magnetic field is governed by the canonical structure
$\omega_{\rm symp} = - {\half} F_{ij} dx^i dx^j$.
We see that the present case where we include the band structure is equivalent to using fermions in free space
with a modified field, namely $\left( F - \Omega^{-1} \right)_{ij}$ in place of $F$.
The phase volume corresponding to this $\omega_{\rm symp}$ is given by
\beq
d\mu = { 1\over k!} \left( {\Omega \over 2\pi} \right)^k \, { 1\over k!} \left( {{F-\Omega^{-1}} \over 2\pi} \right)^k
\label{band10}
\eeq
This shows that we also have an overall factor of $\left(\Omega \right)^k$ with
integration over all momenta; this factor will take the place of the
filling fraction. 

The effective action we obtained was the Chern-Simons action
associated with the Dolbeault index density, considering the
Landau problem as fermions in free space coupled to the magnetic field.
In the light of the modified structure (\ref{band9}) and
(\ref{band10}), we
see that we can transcribe the action to the case of interest (with
fermions drawn from energy bands) by making two changes, namely,
making the replacement $F \rightarrow F - \Omega^{-1}$
and including an integration over all momenta with the density
${ 1\over k!} \left( {\Omega \over 2\pi} \right)^k$. Thus if $S_{\rm eff}$ denotes the Chern-Simons action
obtained from the Dolbeault index density as given in \cite{KN}, the
action for the case of fermions in electronic bands will be
\beq
S_{\rm eff/band} = \int_{BZ} \,{ 1\over k!} \left( {\Omega \over 2\pi} \right)^k
S_{\rm eff} ( F\rightarrow F - \Omega^{-1})
\label{band11}
\eeq
The integration over the momenta is over the Brillouin zone of momentum states for the energy band.
From this effective action one can easily extract a general expression for the Hall currents for a fully filled band with Abelian gauge fields. By taking the variation with respect to $A_\mu$, we get
\beqar
*J &=& \int_{BZ} \,{ 1\over k!} \left( {\Omega \over 2\pi} \right)^k \Bigl[ ~
\I ~\Bigr]_{F \rightarrow F-\Omega^{-1}}\nonumber\\
J^\mu &=&  \e^{\mu \a_1 \a_2 \cdots \a_{2k}}\int_{BZ} \,{ 1\over k!} \left( {\Omega \over 2\pi} \right)^k \Bigl[ ~
\I_{\a_1 \a_2 \cdots \a_{2k}} ~\Bigr]_{F \rightarrow F-\Omega^{-1}}
\label{current}
\eeqar
where $\I$ is given in (\ref{SLLL}).

This is the general expression, but it may be somewhat cryptic for
straightforward application. Therefore, we will
first consider $S_{\rm eff}$ given in
(\ref{band19}) to (\ref{band21}) for the case of 2+1, 4+1 and 6+1 dimensions
and work out the Hall currents.
The general formula valid for all dimensions and with nonzero
background curvatures
will be worked out at the end of this section.

\noindent\underline{(2+1) dimensions}

\beq
J^i =  \epsilon^{ij} \left( {E_j \over 2\pi} + {1 \over 2} { R_{j0} \over 2\pi}\right)~\nu_1
\label{current2}
\eeq
where $E_i = F_{i0}$ and $\nu_1$ is the integral of the first Chern class, given by
\beq
\nu_1 = \int_{BZ} {\Omega \over 2\pi} 
\label{CC1}
\eeq
The curvature $R_{j 0}$ takes values in $U(1)$ as explained
in the introduction.

Expression (\ref{current2}) agrees with previous results in the case of flat 2d spaces. An interesting feature of (\ref{current2}), which is also valid in all higher dimensions, is that {\it a Hall current can be generated from time variation of the metric even if there is no external electric field applied to the system}. While we make a note of this interesting fact, in
deriving the Hall current in higher dimensions we will neglect this effect
for simplicity and consider manifolds whose curvature is time-independent. 

\noindent\underline{(4+1) dimensions}

The relevant expression for $\I$ in (4+1) dimensions is
\beq
\I = {1 \over 2} {F \over 2\pi} \wedge {F \over 2\pi} + {F \over 2\pi} \wedge {c_1 \over 2} + {1 \over 12} (c_1^2 + c_2)
\label{index4} 
\eeq
where $c_1,~c_2$ are given in (\ref{band16a}).
Making the substitution $F \rightarrow F-\Omega^{-1}$ and integrating over the momentum space with density ${ 1\over k!} \left( {\Omega \over 2\pi} \right)^k$, as explained above in equation (\ref{current}), produces the following expression for the electromagnetic Hall current
\beq
J^i = {1 \over 2} {1 \over {(2\pi)^2}} \epsilon^{ijkl} E_j \left( F_{kl} + {\Tr\, R_{kl} \over 2} \right)~\nu_2 + {1 \over 2\pi} E_j \nu_1^{ij} 
\label{current4}
\eeq
where $\nu_k$ and $\nu_1^{ij}$ are given by
\beqar
\nu_k &=& \int_{BZ} {1 \over k!} \left( {\Omega \over 2\pi} \right)^k \label{nCC} \\
\nu_1^{ij} &=& \int_{BZ} {\Omega^{ij} \over {(2\pi)^{2 k-1}}} d^{2k} p
\label{Csub}
\eeqar
Here $\nu_k$ in (\ref{nCC}) is the integral of the k-th Chern class
over the band of electronic states.
In deriving (\ref{current4}) we used the relation
\beq
\left(\Omega^{-1} \right)_{ij} \epsilon_{\alpha_1\alpha_2 \cdots \alpha_{2k}} ~\Omega^{\alpha_1\alpha_2} \cdots \Omega^{\alpha_{2k-1}\alpha_{2k} } = -2 k ~\epsilon_{ij \alpha_1 \cdots \alpha_{2k-2}}\Omega^{\alpha_1\alpha_2} \cdots \Omega^{\alpha_{2k-3}\alpha_{2k-2} }
\label{sub1}
\eeq
The expression (\ref{current4}) for the current agrees with the expression derived in \cite{currents} for flat manifolds, but now includes
generalization to curved manifolds, in addition to the virtue of being derived purely from a topological point of view.

\noindent\underline{(6+1) dimensions}

The relevant index density in (6+1) dimensions is
\beq
{\rm index~density} = {1 \over 3!} \left({F \over 2\pi}\right)^3 + {1 \over 2} \left({F \over 2\pi}\right)^2 \wedge {c_1 \over 2} + {F \over 2\pi} \wedge {1 \over 12} (c_1^2 + c_2) + {1 \over 24} (c_1 c_2)
\label{index6} 
\eeq
Using (\ref{current}) we obtain the following expression for the electromagnetic Hall current,
\beqar
J^i &=& \epsilon^{ijklrs} {1 \over {2^3(2\pi)^3}} E_j \left[ \left(F_{kl} + {1 \over 2} \Tr R_{kl} \right) ~ \left(F_{rs} + {1 \over 2} \Tr R_{rs} \right) ~-~{1 \over 12} \Tr\left( R_{kl}R_{rs} \right) \right] \nu_3 \nonumber \\ && +  {1 \over {2(2\pi)^2}}~E_j \left(F_{kl} + {1 \over 2} \Tr R_{kl} \right) \nu_2^{ijkl} ~+~{1 \over {2\pi}} E_j \nu_1^{ij} 
\label{current6}
\eeqar
where $\nu_3$ is the integral of the 3rd Chern class, $\nu_1^{ij}$ is defined in (\ref{Csub}) and $\nu_2^{ijkl}$ is given by
\beq
\nu_2^{ijkl} = \int {{\Omega^{ij}\Omega^{kl} + \Omega^{il}\Omega^{jk} + \Omega^{ik}\Omega^{lj}} \over {(2\pi)^4}}~d^{6} p
\label{Csuba}
\eeq
In deriving (\ref{current6}) we have used a relation similar to (\ref{sub1}).
The general identity is given in (\ref{AppB8}) in Appendix B.
The specific cases we need here are
\beqar
\nu_{1}^{ij}&=&\int_{BZ}{1\over 3!} \left( {\Omega \over 2 \pi}\right)^3 \e^{ij\a_1\a_2\a_3\a_4}
{1\over 2!}{(\Omega^{-1})^{\a_1\a_2} \over 2 (2\pi)}{(\Omega^{-1})^{\a_3\a_4} \over 2 (2\pi)}\nonumber\\
 \nu_2^{ijkl}&=& - \int_{BZ} {1\over 3!} \left( {\Omega \over 2 \pi}\right)^3 \e^{ijkl ab}
{(\Omega^{-1})^{ab} \over 2 (2\pi)}
\label{sub3}
\eeqar
Again the expression (\ref{current6}) for the Hall current agrees with previous results derived for flat backgrounds \cite{currents} and further generalizes them to curved manifolds. 

\noindent\underline{$(2 k+1)$ dimensions}

We will now write down the general expression for the Hall current for an arbitrary $2k$-dimensional (complex) curved manifold. 
\beqar
J^i &&= \epsilon^{iji_1\cdots i_{2k-2}} \sum_{s=0}^{k-1} {{E_j \left(F^{k-s-1} \right)_{i_1i_2 \cdots i_{(2k-2s-2)}}} \over {(2 \pi)^k 2^{k-1} (k-s-1)!}}  \left[ ({\rm td})_s \right]_{i_{(2k-2s-1)} \cdots i_{(2k-2)}} ~\nu_k \label{current2k}\\
+ &&\sum_{s=0}^{k-1} \sum_{l=1}^{k-s-1} { {E_j \left(F^{k-s-l-1} \right)_{i_1i_2 \cdots i_{(2k-2s-2l-2)}}}\over {(2 \pi)^{k-l} 2^{k-l-1} (k-s-l-1)!}} \left[ ({\rm td})_s \right]_{i_{(2k-2s-2l-1)}  \cdots i_{(2k-2l-2)}} ~\nu_{k-l}^{iji_1\cdots i_{(2k-2l-2)}} \nonumber
\eeqar
where $\left[ {\rm td} \right]_s$ is the $2s$-form in terms of curvatures in the expansion of the Todd class as given in (\ref{band14}). $\nu_k$ is the integral of the k-th Chern class defined in (\ref{nCC}) and the integrals $\nu_{k-l}$ over the sub-classes  are defined by
\beq
\nu_{k-l}^{iji_1\cdots i_{(2k-2l-2)}} = \int { d^{2k} p \over {(2\pi)^{k+l}}}  ~\sum_{\rm dist. perm.} \Bigl[ \Omega^{ij}\Omega^{i_1i_2} \cdots \Omega^{i_{(2k-2l-3)}i_{(2k-2l-2)}}
 \Bigr]
\label{subclassn}
\eeq
The differential $2s$-form corresponding to the Todd class is written as
\beq
({\rm td})_s = {1\over 2^s (2\pi)^s} \,[({\rm td})_s]_{a_1 \cdots a_{2s}}  (dx^{a_1} \cdots
dx^{a_{2s}} )
\label{todd-corr}
\eeq
so that $[({\rm td})_s]_{a_1 \cdots a_{2s}} $ is given in terms of traces of powers of the curvature with indices as shown. Factors of $2$ and $2\pi$ have been
separately included in (\ref{current2k}).
\section{Hall viscosity in higher dimensions}
The effective actions we have derived in \cite{KN}  also allow for the direct evaluation of the Hall viscosity beyond two dimensions. Here we shall illustrate in detail the derivation of the Hall viscosity in the case of the two and four dimensional QHE but similar arguments apply to all 
dimensions. (For the viscosity, we will consider QHE
with electrons in free space.)
 Of course in the case of the two-dimensional QHE our results agree with previous results derived in \cite{visc}-\cite{AG2}.

We start by recalling that viscosity is defined in terms of the two-point function
$\la T^{\mu\nu}(x) T^{\rho\sigma}(y) \ra$ for the energy-momentum tensor $T^{\mu\nu}$. We can identify the viscosity
by considering the expansion
of $T^{\mu\nu}$ (obtained from the effective action)
in terms of powers of derivatives of the metric.
The term involving the time-derivative of the metric
gives the viscosity. In mathematical terms
\beq
T^{\mu\nu} = \eta^{\mu\nu}_{~\,\rho\sigma}  ~ {\dot g}^{\rho\sigma} + \cdots
\label{vis1}
\eeq
and we identify $\eta^{\mu\nu}_{~\rho\sigma}$ with the viscosity tensor.
Since $T^{\mu\nu}$ is obtained from the variation of the action with respect to the (inverse) metric $g^{\mu\nu}$, we can also write this as
\beq
\la T^{\mu\nu}(x) T_{\rho\sigma}(y) \ra 
= 4 g^{\mu\lambda} g^{\nu\tau} {\delta^2 S_{\rm eff} \over \delta g^{\rho\sigma}(y) \delta g^{\lambda\tau}(x)}
= \eta^{\mu\nu}_{~\, \rho\sigma} {\del \over \del x^0} \,\delta^{(2k+1)} (x, y)
+ \cdots
\label{vis1a}
\eeq
This agrees with the usual definition in terms of
the two-point correlation function for the energy-momentum tensor.

In using (\ref{vis1}) for calculating the Hall viscosity, we should keep in mind that
the effective action was obtained for complex manifolds. A general variation of the metric, which does not necessarily preserve the complex structure, is needed for the correlation function in (\ref{vis1}) or (\ref{vis1a}). Of course, one can, after identifying
$\eta^{\mu\nu}_{~\rho\sigma}$, set the background metric to its value appropriate for the complex manifold of interest. 
The spin connection involves the $U(k)$ subalgebra of the
vector representation of the Lie algebra of
the $SO(2 k)$ holonomy group.
Therefore, to carry out a general variation of the metric,
we need the relation between the complex 
$U(k)$ spin connection ($\omega^0,~\omega^a)$ and the corresponding
real $SO(2k)$ quantities ${\omega}^{\a\b}$. This is worked out in detail in
Appendix A for the two and four dimensional cases. 

The $SO(2k)$ spin connection ${\omega}$ is related to the Christoffel symbols 
$\Gamma^\alpha_{\mu\,\beta}$ by\footnote{Our conventions, given in Appendix A, are that Greek letters from the beginning of the alphabet
denote tangent frame indices, lower case Roman letters indicate
spatial components in the coordinate basis, and Greek letters from later in the alphabet denote coordinate basis again, but including space and time
components.}
\beq
{\omega}_\mu^{\a\b} = e^{\a}_i \, \Gamma^i_{\mu j} (e^{-1})^{j\b} - \del_\mu e^{\a}_j\,(e^{-1})^{j\b} 
\label{vis8}
\eeq
In some formulae we will use the form notation
${\omega}^{\a\b} = {\omega}^{\a\b}_{\mu} \,dx^\mu$
and $  \Gamma^i_{j}  = \Gamma^i_{\mu\,j} \,dx^\mu$
where $e^{\a}_i$ is the frame field in general. Since we deal with a nonrelativistic system we also have the specific values
\beq
g_{00}=1,~~~g_{0i}=0,~~~e^{0}_{i}=e^{i}_0=0
\eeq
The variation of the expression (\ref{vis8}) gives
\beqar
\delta {\omega}^{\a\b} &=& e^{\a}_i \left( \delta \Gamma^i_j 
\right) (e^{-1})^{j\b} - \left( d \Theta + {\omega}\, \Theta - \Theta\, {\omega} \right)^{\a\b}\nonumber\\
\delta \Gamma^i_{\mu\,j} &=&
{1\over 2} g^{il} \left( - \nabla_l \delta g_{\mu j}
+ \nabla_\mu \delta g_{j l} + \nabla_j \delta g_{\mu l}
\right)\label{vis9}\\
\Theta &=& \delta  e^{\a}_i\,(e^{-1})^{i\b}
\nonumber
\eeqar
\subsection{Two dimensional QHE}
The topological bulk effective action in two dimensions for QHE for the $s$-th Landau level is given by
\beq
S_{2+1}^{( s)} = {1 \over {4\pi}}  \int \left[ A dA +2 \bar{s} A d\omega^0 +   \left( \bar{s}^2 - {1\over 12} \right)  \omega^0 \, d \omega^0  \right]
\label{eff2d}
\eeq
where we defined $\bar{s} = s+ \half$ and we switched to a hermitian basis for the gauge and spin connections
which eliminates the factor of $i^2$ from (\ref{band19}).
Further, as explained in Appendix A, 
\beq
\omega^0 = \half \epsilon^{\a\b}{\omega}^{\a\b}, \hskip .3in
\omega^0 d \omega^0 = -\half ({\omega^{\alpha\beta}}d{\omega^{\beta\alpha}})
\label{100}
\eeq
Varying $\omega^{\alpha\beta}$, we get
\beq
\delta S^{(s)}_{\rm 3d}
= {1 \over 4 \pi} \int \bar{s}\, dA ~\epsilon^{\a\b}  \delta {\omega}^{\a\b}   -  \left( \bar{s}^2 - {1\over 12} \right) (\delta {\omega^{\alpha\beta}}\, d {\omega^{\beta\alpha}} )
\label{101}
\eeq
We will use (\ref{vis9}) and evaluate the two terms in (\ref{101})
separately. For the first one, we find
\beqar
\int dA ~\epsilon^{\a\b} ~\delta {\omega}^{\a\b} &=&  \int dA ~\epsilon^{\a\b} \left[ e \delta \Gamma e^{-1} -d \Theta - [{\omega} , \Theta] \right]^{\a\b} 
\nonumber \\
&=& {1 \over 2} \int \epsilon^{\a\b} e^{\a}_i e^{-1 j \b} ~\delta \Gamma^{i}_{\mu j} ~\epsilon^{\mu \nu \sigma} ~F_{\nu \sigma} ~d^3x 
\label{102}
\eeqar
In going from the first to the second line of (\ref{102}) we used the fact that
 $\epsilon^{\a\b} ~[{\omega}~,~\Theta]^{\a\b}=0$ in two dimensions. We can evaluate the second term in (\ref{101}) in a similar way:
\beqar
\int (\delta {\omega^{\alpha\beta}}\, d {\omega^{\beta\alpha}} ) &=& {1 \over 2} \int  e^{\a}_i e^{-1 j \b} ~\delta \Gamma^{i}_{\mu j} ~\epsilon^{\mu \nu \sigma} ~({R}_{\nu\sigma})^{\b\a} ~d^3x \nonumber \\
&=& {1 \over 2} \int \delta \Gamma^{i}_{\mu j} ~\epsilon^{\mu \nu \sigma} 
~({R}_{\nu \sigma})^j_{~i} ~d^3x
\label{103}
\eeqar
The curvatures in (\ref{102}), (\ref{103}) are defined by
\beq 
F_{\nu\sigma} = \del_\nu A_\sigma - \del_\sigma A_\nu, \hskip .3in
({R}_{\nu\sigma})^j_{~i} =
(e^{-1})^{j \b} \, e^{\a}_i  \,
(\del_\nu \omega_\sigma^{\b \a} - 
\del_\sigma \omega_\nu^{\b \a} ) 
\label{104}
\eeq
Using $\delta \Gamma$ as given in (\ref{vis9}) we find that the variation of the effective action (\ref{101}) becomes
\begin{align}
\delta S^{(s)}_{\rm 3d} =& {1 \over 16\pi} \int \left( 
- \nabla_l \delta g_{\mu j} + \nabla_\mu \delta g_{l j}
+ \nabla_j \delta g_{\mu l}\right)\nonumber\\
&\hskip .5in \times \left[ e^{-1 l \a} e^{-1j \b} \epsilon^{\a\b} \bar{s} F_{\nu \sigma}  - \left(\bar{s}^2- {1 \over 12}\right) ({R}_{\nu \sigma})^{jl} \right]\e^{\mu\nu\sigma} ~d^3x\nonumber \\
=& -{1 \over 8\pi} \int \delta g_{\mu l} \left[ \bar{s}~(J^0)^{lj} ~\nabla_j F_{\nu\sigma}  - \left(\bar{s}^2- {1 \over 12}\right) \nabla_j ({R}_{\nu\sigma})^{jl} \right] \e^{\mu\nu\sigma} ~d^3 x
\label{105}
\end{align}
In obtaining the last line of (\ref{105}) we have done an
integration by parts and also defined the antisymmetric tensor
\beq
(J^0)^{lj} = e^{-1 l \a} e^{-1j \b} \epsilon^{\a\b}
\label{106}
\eeq
The energy-momentum tensor can be read off from the variation
$\delta S_{\rm eff}$ using the usual formula,
\beq
\delta S_{\rm eff} = - {1 \over 2} \int \delta g_{ml} T^{ml} \sqrt{\det g} ~d^n x
\label{108}
\eeq
Comparing (\ref{105}) and (\ref{108}) we identify the energy-momentum tensor for the 2d QHE as
\beq
T^{m l} =   -{1 \over {4 \pi\sqrt{\det g}}} \left( g^{i l} \e^{mk} + 
g^{i m} \e^{lk} \right)
\left[ \bar{s} (J^0)_i~^j   \nabla_j F_{0k} 
- \left(\bar{s}^2 -{1 \over 12}\right) \nabla_j (R_{0 k})^j_{~i} \right]
\label{109}
\eeq
In order to calculate the Hall viscosity we need to identify the terms 
in this expression which are proportional to the time-derivative of the metric. For the covariant derivative of $F$ we can use
\beqar
\nabla_j F_{0k} 
&=& \del_j F_{0k} - \Gamma_{j 0}^n F_{n k} 
- \Gamma^n_{jk} F_{0n} 
\nonumber\\
&=&- {1\over 2} g^{ n l } {\dot g}_{l j}\, F_{n k} + \cdots
\label{110}
\eeqar
where the ellipsis indicates terms that do not contain ${\dot g}$. 
As for the curvature term, we find
\beq
(R_{0k})^j_{~i} = {1\over 2} g^{j n} \left( \nabla_i {\dot g}_{nk} - \nabla_n {\dot g}_{ik}\right)
\label{111}
\eeq
Therefore
\beq
\nabla_j (R_{0k})^j_{~i} 
= {1\over 2}   \nabla_j \nabla_i \left(g^{jn}{\dot g}_{nk} \right)- {1 \over 2} g^{jn} \nabla_j \nabla_n {\dot g}_{ik}
- {1\over 2} g^{nl} {\dot g}_{lj}\, ({R}_{nk})^j_{~i}
\label{112}
\eeq
We can simplify this result further by commuting the covariant derivatives in the first term and writing 
\beq
\nabla_j \nabla_i \left(g^{jn}{\dot g}_{nk} \right) = \nabla_i \nabla_j \left(g^{jn}{\dot g}_{nk} \right) + (R_{ji})^j_{~m}~\left(g^{mn}{\dot g}_{nk} \right) - ({R}_{ji})_k^{~m}~\left(g^{jn}{\dot g}_{nm} \right)
\label{113}
\eeq
Using (\ref{113}) in (\ref{112}) we get
\beqar
\nabla_j (R_{0k})^j_{~i} 
&=& {1\over 2}   \nabla_i \nabla_j \left(g^{jn}{\dot g}_{nk} \right)- {1 \over 2} \nabla^2 {\dot g}_{ik}
- {1\over 2} g^{nl} {\dot g}_{lj}\, ({R}_{nk})^j_{~i}\nonumber\\
&&\hskip .2in  +{1 \over 2} g^{mn}{\dot g}_{nk} ~({R}_{ji})^j_{~m}~ - {1 \over 2} g^{jn}{\dot g}_{nm}~({R}_{ji})_k^{~m}~
\label{114}
\eeqar
An arbitrary perturbation of the metric is rather too general for our purpose, since some of it corresponds simply to a coordinate change or diffeomorphism. A suitable covariant gauge choice which restricts 
the variations appropriately is the de Donder gauge which is
given by \cite{donder}
\beq
\nabla_j (g^{j n} {\dot g}_{n k})
- {1\over 2} \nabla_k (g^{j r} {\dot g}_{jr})
= 0
\label{115}
\eeq
It is possible to choose such a gauge for the perturbations of the metric
by using the freedom of coordinate transformations. 

Using (\ref{110}), (\ref{114}), (\ref{115}) in the expression (\ref{109}) for the energy-momentum tensor we find that the term linear in ${\dot g}$ is of the form
\beqar
T^{m l} &=&   {1 \over {8 \pi\sqrt{\det g}}} \left( g^{i l} \e^{mk} + 
g^{i m} \e^{lk} \right)
\biggl[ \left( \bar{s} (J^0)_i~^j   g^{rs} F_{sk}\right) {\dot g}_{rj} 
+ \left(\bar{s}^2 -{1 \over 12}\right) \Bigl( {1 \over 2} \nabla_i \nabla_k (g^{rn} {\dot g}_{rn}) \nonumber \\
&&-  \nabla^2 {\dot g}_{ik}
-  g^{nl} {\dot g}_{lj}\, ({R}_{nk})^j_{~i} + g^{mn}{\dot g}_{nk} ~({R}_{ji})^j_{~m}~ - g^{jn}{\dot g}_{nm}~({R}_{ji})_k^{~m} \Bigr)\biggr]
\label{116}
\eeqar

In two dimensions there are further simplifications since the Riemann tensor has the form
\beq
R_{ijkl} = {R \over 2}
\left( g_{ik} g_{jl} - g_{il}\,
g_{jk} \right)
\label{117}
\eeq
where $R$ is the Ricci scalar curvature. Further 
\beq
(J^0)^{lj} = e^{-1 l \a} e^{-1j \b} \epsilon^{\a\b}={ \epsilon^{lj} \over \sqrt{\det g}}
\label{118}
\eeq
By use of these expressions, the result
(\ref{116}) for the energy-momentum tensor can be simplified as
\begin{align}
T^{m l} =   {1 \over {8 \pi\sqrt{\det g}}} \left( g^{mi} \e^{lk}+ g^{li} \e^{mk}  
 \right)
\biggl\{&\left[ \bar{s} B   
+ \left(\bar{s}^2 -{1 \over 12}\right) \left( {R \over 2} -\nabla^2 \right)\right]  {\dot g}_{ki}
 \nonumber \\
&+ {1 \over 2} \left(\bar{s}^2 -{1 \over 12}\right) \nabla_i \nabla_k \left(g^{rn} {\dot g}_{rn} \right) \biggr\}
\label{119}
\end{align}
In this expression, we used the
magnetic field $B$ given by
\beq
F_{ij} = \epsilon_{ij} ~B~\sqrt{\det g}
\label{120}
\eeq
Comparing (\ref{119}) with the expression (\ref{vis1}) of the energy-momentum tensor in terms of the Hall viscosity, we see that we can write 
\beqar
\sqrt{\det g}~T^{m l} &=&   {1\over 2} \, \eta_{H} \, \left( g^{mi} \e^{lk}+ g^{li} \e^{mk}  
 \right)  {\dot g}_{ki}
 \nonumber \\
&&+ {1 \over 2}\, \eta^{(2)}_{H} \,\left( g^{mi} \e^{lk}+ g^{li} \e^{mk}  
 \right) \nabla_i \nabla_k \left(g^{rn} {\dot g}_{rn} \right) 
\label{120}
\eeqar
where the coefficients can be read off as
\beqar
\eta_H &=& {1 \over 4\pi} \left[ \bar{s} B + \left(\bar{s}^2 -{1 \over 12}\right) \left( {R \over 2} + \vec{k}^2 \right) \right]\nonumber \\
\eta^{(2)}_{H} & =& { 1 \over 8\pi} \left(\bar{s}^2 -{1 \over 12}\right)
\label{121}
\eeqar
The magnetic field and curvature dependent terms of $\eta_H$, and the structure of 
$T^{m l} $ as in (\ref{120}), are in agreement with
\cite{AG1}. Notice that the coefficient of ${\dot g}_{rs}$ in
(\ref{119}) and (\ref{120}) is an operator, so an expansion in terms of the eigenmodes of the covariant Laplacian will be needed to identify numerical values.
For purposes of comparison, we have indicated the eigenvalue of $-\nabla^2$
as $\vk^2$, which would be appropriate in the flat space limit.
\subsection{ Four dimensional QHE}
Turning to 4+1 dimensions, we write the action (\ref{band20}) as
\beqar
S_{4+1}^{(s)} &=&{ (s+1)\over {(2\pi)^2}} \int \Biggl\{ { 1 \over 3!} \Bigl(A+(s+1)\omega^0\Bigr) \Bigl[d\Bigl(A+(s+1)\omega^0\Bigr)\Bigr]^2 \nonumber \\
&&-{ 1 \over 12} \Bigl(A+(s+1)\omega^0\Bigr) \Biggl[  (d\omega^0)^2 + {1 \over 4} {R^a} \wedge { R^a}-s\left({s \over 2}+1\right) {R^a} \wedge { R^a} \Biggr] \Biggr\}
\label{122}
\eeqar
where we have removed the overall factor of $i^3$ by going over to the hermitian forms of the connections. We have also used
$j = s/2$.
The relation between the $U(2)$ spin connections and curvatures used in (\ref{122}) and the corresponding $SO(4)$ quantities is derived in (\ref{20A}) and is given by
\beqar
\omega^0 &=& {1\over 4} \epsilon^{\a\b} {\omega}^{\a\b},
\hskip .2in R^0 = {1\over 4} \epsilon^{\a\b} d{\omega}^{\a\b} \nonumber \\
R^a R^a &=&- 4 R^0R^0 - {R}^{\a\b} {R}^{\b\a}
\label{123}
\eeqar 
Using these relations, the effective action (\ref{122}) can be expressed as
\begin{align}
S_{4+1}^{(s)} = { (s+1)\over {(2\pi)^2}}\int  \biggl[&
{ 1 \over 3!}   \A  d\A d\A  -  {1\over 12} \int  \A \Bigl[ 4s \left({s \over 2} +1\right) d\omega^0 d\omega^0\nonumber\\
&
+ \left( s\left( {s \over 2} +1\right) - {1 \over 4}\right) {R}^{\a\b} \wedge {R}^{\b\a} \Bigr] \biggr]
\label{124}
\end{align}
where $\A = A + (s+1) \omega^0$. 

The variation of the effective action (\ref{124}) naturally splits into
two types of terms of the form
\begin{align}
\delta S = & ~\delta S^{(1)} + \delta S^{(2)}
\label{128}\\
\delta S^{(1)} = &~ { (s+1)\over {(2\pi)^2}}\int  \delta  \omega^0\, K
\label{129}\\
\delta S^{(2)} = & -{ (s+1)\over {12 (2\pi)^2}} \left[s \left({s \over 2} +1\right) - { 1 \over 4}\right] \int \A\, \delta (R^{\a\b} R^{\b\a})
\label{131}\\
K = &~{s+1 \over 2} dAdA + {2 \over 3} \left[(s+1)^2 + {1 \over 2}\right] dA d\omega^0 \nonumber\\
&+{s+1 \over 2} d\omega ^0 d\omega ^0 - {s+1 \over 12} \left[s \left({s \over 2} +1\right) - { 1 \over 4}\right] R^{\a\b} R^{\b\a}
\label{130}
\end{align}
The variation and simplification of these terms
will proceed along lines similar to the $(2+1)$-dimensional case.
Using (\ref{vis9}) and an integration by parts for the
$d \Theta$-term, we find
\beq
 \int \delta \omega^0\, K  =  {1 \over 4} \int 
\biggl[ (J^0)^{l j} \nabla_j \delta g_{\mu l}  -\left(\delta e e^{-1}\right)^{\a\b} ([ \omega_\mu, \epsilon])^{\b\a} 
  \biggr]
K^\mu \sqrt{\det g} 
\label{125}
\eeq
where
\begin{align}
K^\mu \sqrt{\det g} = \e^{\mu\nu\sigma \tau \rho} \biggl\{ & {s+1 \over 2} \del_\nu A_{\sigma}\, \del_\tau A_{\rho} + {2 \over 3} \left[(s+1)^2 + {1 \over 2}\right] \del_\nu A_{\sigma}\, \del_\tau \omega^0_{\rho}\nonumber\\
& +{s+1 \over 2} \del_\nu \omega ^0_{\sigma}\, \del_\tau\omega^0_{\rho}  \nonumber \\
&- {s+1 \over 48} \left[s \left({s \over 2} +1\right) - { 1 \over 4}\right] R_{\nu\sigma}^{\a\b}~ R_{\tau\rho}^{\b\a} \biggr\}
\label{126}
\end{align}
and $(J^0)^{lj}$ is given in (\ref{106}). 
The contribution of the second term in (\ref{125}) to the symmetrized version of the energy-momentum tensor is zero. 
(The variation of the frame field as in 
$\left(\delta e e^{-1}\right)^{\a\b}$ can be related to 
the variation of the metric (which is the symmetric combination)
and an antisymmetric part. It is the symmetric part which is relevant for
the energy-momentum tensor.\footnote{The antisymmetric part can be related to spin densities and can be relevant for some other transport coefficient related to the correlation function for the energy-momentum tensor and the spin density. This is not our focus at this stage.}) With an integration by parts, 
(\ref{129}) simplifies as
\beq
\delta S^{(1)} =  - {s+1\over {8 (2\pi)^2}} \int \delta g_{m l}  
 \nabla_j \left[ (J^0)^{lj} K^m + 
(J^0)^{m j} K^l \right] 
\,\sqrt{\det g} 
\label{127}
\eeq
Comparing this with (\ref{108}) we find that the contribution to the energy momentum tensor from (\ref{127}) is
\beq
(T^{ml})^{(1)} = { (s +1) \over {4(2\pi)^2}} 
\nabla_j \left[ (J^0)^{lj} K^m
+ (J^0)^{mj} K^l \right]
\label{132}
\eeq

For the evaluation of the second type of terms, namely,
(\ref{131}), we notice that
\beq
\delta ~\tr (R\wedge R) = d ~\tr ~ \delta (\omega d \omega + {2 \over 3} \omega^3) = 2 ~d ~\tr (\delta \omega R)
\label{133}
\eeq
(The trace here, indicated by $\tr$, is for $\omega$, $R$ in the vector representation of
$SO(2k)$.)
Using (\ref{133}) and (\ref{vis9}) in (\ref{131}) we then get
\beqar
\delta S^{(2)} = -{ (s+1)\over {6 (2\pi)^2}} \left[s \left({s \over 2} +1\right) - { 1 \over 4}\right] \int d \A ~ \tr \left( \delta \Gamma e^{-1} R e + \delta e e^{-1} \left( dR + [\omega , R]\right) \right]
\label{134}
\eeqar
The last term in (\ref{134}) vanishes by the Bianchi identity $dR + [\omega , R]=0$. After writing $\delta \Gamma$ in terms of variations of the metric and carrying out a partial integration we find 
\beq
\delta S^{(2)} = { (s+1)\over {12 (2\pi)^2}} \left[s \left({s \over 2} +1\right) - { 1 \over 4}\right] \int \delta g_{\mu l}   \nabla_j \left[ (R_{\nu \sigma})^{jl} \del_\tau \A_\rho\right] \e^{\mu\nu\sigma \tau \rho}~d^5 x
\label{135}
\eeq
Comparing this with (\ref{108}) we find that the contribution to the energy-momentum tensor from (\ref{135}) is
\beq
(T^{ml})^{(2)} = - { (s+1)\over {12 (2\pi)^2}} \left[s \left({s \over 2} +1\right) - { 1 \over 4}\right] \left[ \nabla_j \left[ (R_{\nu \sigma})^{j l} \del_\tau \A_\rho\right] { \e^{m\nu\sigma \tau \rho} \over \sqrt{\det g}} + (m \leftrightarrow l) \right]
\label{136}
\eeq

In order to identify the Hall viscosity we have to extract terms linear in ${\dot g}$ in (\ref{132}) and (\ref{136}). To simplify the calculation we will neglect terms of the form $\del {\dot g}$ which will produce momentum-dependent terms for the Hall viscosity.

Focusing first on $(T^{ml})^{(1)}$ and using the fact that
\beq
\nabla_\mu  (J^0)^{lj} = e^{-1 l \a} [ \epsilon ~, ~ \omega_\mu ] ^{\a\b} 
e^{-1 j \b }
\label{137}
\eeq
we find
\beq
\nabla_j \left( (J^0)^{lj} K^m \right) = e^{-1 l \a}e^{-1 j \b } \left( [ \epsilon ~, ~ \omega_j ]^{\a\b} K^m +  \epsilon^{\a\b} \del_j K^m + \epsilon^{\a\b} \Gamma^m_{j \rho} K^{\rho} \right)
\label{138}
\eeq
The first term in (\ref{138}) vanishes if $\omega$ preserves the $U(2)$ structure as expressed in (\ref{relation}). For a constant magnetic field the second term in (\ref{138}) will contribute only momentum-dependent terms, 
of the form $\del {\dot g}$, to the expression for the Hall viscosity. The third term leads to 
\beq
(T^{ml})^{(1)} = {{s+1} \over {8 (2\pi)^2}} \left(g^{mn} (J^0)^{lj} + g^{ln} (J^0)^{mj} \right)~{\dot g}_{nj} K^0 
\label{137a}
\eeq
Turning to $(T^{ml})^{(2)} $, we notice that 
\beq
{\tilde K}^{j l m} = \left[ (R_{\nu \sigma})^{j l} \del_\tau \A_\rho\right] { \e^{m\nu\sigma \tau \rho} \over \sqrt{\det g}}
\label{137b}
\eeq
transforms as a rank 3 contravariant tensor, so its covariant derivative is easy to write down. If any of the indices $\nu, \sigma, \tau, \rho$ 
is taken to be the time-component, the corresponding contribution will
involve covariant derivatives of ${\dot g}$. Since we are not including them here, the only contribution is from $\Gamma^m_{j 0} {\tilde K}^{j l 0}$
in the expression for the covariant derivative. 
So the term linear in ${\dot g}$ in (\ref{136})(without covariant derivatives on it) is of the form
\beq
(T^{ml})^{(2)} =  { (s+1)\over {24 (2\pi)^2}} \left[s \left({s \over 2} +1\right) - { 1 \over 4}\right] 
\left(g^{mn}  (R_{rs})^{lj}+ g^{ln} (R_{rs})^{mj} \right)~{\dot g}_{nj}
\del_p \A_q { \e^{rspq} \over \sqrt{\det g}}
\label{138a}
\eeq
These two expressions, namely (\ref{137a}) and (\ref{138a}),
give us the momentum-independent terms of the Hall viscosity.
To simplify further, we will consider it in 
two particular limits: 1) The flat limit where the 
$\mathbb{CP}^2$ radius becomes very large and the curvature vanishes and 2) On the manifold $\mathbb{CP}^2$ with curvatures
set to the values appropriate to this background. 

In the flat limit, $\mathbb{CP}^2$ space decomposes into 
$\mathbb{C} \times \mathbb{C}$, corresponding to the
planes $(1,2)$ and $(3,4)$. The flat limit may be the most pertinent case for the current experimental set-ups \cite{{4DQHE},{4DQHE1}}. Each plane carries a constant perpendicular magnetic field $F_{12}= F_{34}= B= n / 2r^2$. Also we can write
$(J^0)^{lj} \rightarrow \epsilon^{ij}$. Since the curvature terms vanish in this limit the contribution from $(T^{ml})^{(2)}$ is zero. The contribution from $(T^{ml})^{(1)}$ is of the form
\beq
T^{ml} = { (s +1)^2 \over {8(2\pi)^2}} \left( g^{mi} \e^{lk}+ g^{li} \e^{mk} \right) {\dot g}_{ki} B^2
\label{139}
\eeq
Comparing with (\ref{120}) we find that the Hall viscosity in this limit is
\beq
\eta_H = {1 \over 4} \left( {(s+1) B \over 2\pi} \right)^2
\label{140}
\eeq
We have not evaluated the momentum-dependent terms, so there is no
result for such terms in the Hall viscosity.

Turning to the Hall viscosity for the $\mathbb{CP}^2$ background,
it is useful to write the expression for $K^0$ and $(T^{ml})^{(2)}$ in terms of differential forms.
\begin{align}
d^4x\,\sqrt{\det\, g}\, K^0
=& {s+1 \over 2} dA dA + {2\over 3} \left[ (s+1)^2 + {\half} \right] dA d\omega^0 +
{s+1\over 2} d\omega^0\, d\omega^0 \nonumber\\
&- {s+1 \over 12} \left[ s \left({s\over 2}+ 1\right) - {1\over 4}\right]
R^{\alpha\beta} R^{\beta\alpha}
\label{140a}
\end{align}
From the relations given in Appendix A,
$dA = F = n\, \Omega_{\rm K}$, $d\omega^0 = R^0 = (3/2) \Omega_{\rm K}$
and
\beq
R^{\alpha\beta} R^{\beta\alpha} = -4 R^0 R^0 - R^a R^a
= - 6 \,\Omega_{\rm K}^2
\label{140b}
\eeq
where $\Omega_{\rm K}$ is the K\"ahler two-form for $\mathbb{CP}^2$.
Using these relations, $K^0$ becomes
\beq
K^0
= \left\{ (s+1) (n/2)^2 + \left[ (s+1)^2 + {1\over 2} \right] (n/2)
+ {s+1\over 8} \left[ (s+1)^2 + {3}\right] \right\}
\label{140c}
\eeq
Similarly, we write $(T^{ml})^{(2)}$ as
\beq
d^4x \sqrt{\det\, g}\, (T^{ml})^{(2)}
= {s +1 \over 12 (2\pi)^2} 
\left[ \left[ s \left( {s\over 2}+1\right) - {1\over 4} \right]
\left( g^{mn} R^{lj} + g^{ln} R^{mj} \right) {\dot g}_{nj}
d\A \right]
\label{140d}
\eeq
A useful relation is to note that $R^a \wedge \Omega_{\rm K} = 0$
as shown in Appendix A. Also
$d\A = dA + (s+1) d\omega^0 =
(n + {3\over 2}(s+1)) \Omega_{\rm K}$. The expression for $(T^{ml})^{(2)}$ then simplifies as
\begin{align} 
(T^{ml})^{(2)} = & {s +1 \over 8 (2\pi)^2} 
\left[ \left[ s \left( {s\over 2}+1\right) - {1\over 4} \right]
\left( {n \over 2} + {3\over 4} (s+1) \right) \right]
 \left( g^{mi} (J^0)^{lk}+ g^{li} (J^0)^{mk} \right) {\dot g}_{ki}
 \label{140e}
\end{align}
Using $K^0$ from (\ref{140c}) in (\ref{137a}) and adding
$(T^{ml})^{(2)} $ from the equation given above, we get
\begin{align}
T^{ml} = & { (s +1) \over {8(2\pi)^2}} \left( g^{mi} (J^0)^{lk}+ g^{li} (J^0)^{mk} \right) {\dot g}_{ki}\times\nonumber\\
 &\left\{ (s+1) \left({n \over 2}\right)^2 +   \left[ {3\over 2}(s+1)^2  -{1\over4} \right]
 \left({n \over 2}\right)
 + {(s+1)\over 2} \left[ (s+1)^2 - {3 \over 8}\right]\right\}
\label{142}
\end{align}
We notice that the tensorial structure of the energy-momentum tensor in (\ref{142}) is appropriately modified, with the metric and the covariant version of $J^0$, for the $\mathbb{CP}^2$ background compared to (\ref{139}) and that the curvature terms do contribute to the overall factor. However the contribution from the curvature terms is relatively negligible in the large $B$ limit as expected. 
\section{Discussion}
In this paper, we have considered some of the transport properties
of quantum Hall systems in arbitrary even spatial dimensions.
The effective action obtained in \cite{KN} provides a uniform approach 
and a convenient 
starting point for this, as transport coefficients can be obtained by
varying this action with respect to the external fields and the metric.
Specifically, we focus on the
 Hall conductivity and the Hall viscosity which are the
transport properties most relevant from the experimental point of view.
Towards this, we first generalized the effective action from \cite{KN}
to take account of the fact that electrons belong to an energy band
in a solid, rather than being in free space. 
We derived an expression for the electromagnetic Hall current 
valid for any even spatial dimension, displaying various terms proportional to integrals of the Chern classes of the Berry curvature of the 
electronic bands. Additionally, our expressions include the contributions due to the spatial curvature.
We expect that these expressions, with or without the spatial curvature,
will be directly relevant for proposed experimental realizations
in higher dimensions \cite{4DQHE, 4DQHE1}.

We have also given explicit expressions for the Hall viscosity
in two and four spatial dimensions, including terms which depend on the
curvature. While the result for two dimensions agrees with previous work on the calculation of responses, it should be emphasized that our approach
places it within a uniform method of derivation.
The results for four dimensions are obviously new.

An important point worth noting is that, in general, there are several additional transport
coefficients or response functions possible.
Already in two spatial dimensions, we see from 
(\ref{current2}) that the second term is of the form
\beq
J^i = {\nu_1 \over 4\pi} \e^{ij}\, R_{j0}
= {\nu_1 \over 8\pi} \e^{ij} \e_{kl} {\nabla^k (g^{ln} {\dot g}_{nj})
\over \sqrt{\det g}}
\label{disc1}
\eeq
This shows that there is a new transport coefficient $\zeta^i_{mn}$ we can define by
\beqar
\la J^i(x) \, T_{mn}(y) \ra &=& \zeta^i_{mn} {1\over \sqrt{\det g}} \del_0 \delta^{(3)}(x-y)\nonumber\\
\zeta^i_{mn} &=&- {\nu_1 \over 8\pi} \e^{ij} (g_{mj}\e_{kn} + g_{nj}\e_{km})
g^{kl} \nabla_l 
\label{disc2}
\eeqar
(As written $\zeta^i_{mn}$ is an operator and must be interpreted in terms of
eigenfunctions of the gradient operator or in terms of Fourier components.)
This transport coefficient exists for the higher dimensional cases as well, although we have not calculated explicit formulae for it.

Higher dimensions also allow for the possibility of nonabelian background gauge fields.
The responses to varying the nonabelian gauge field background
will constitute another set of transport coefficients.
Finally, we have already noted, in the footnote after (\ref{126}),
that one can also have transport coefficients
with correlation functions involving the spin density.
In principle, all such 
additional transport coefficients can be calculated using the effective action
 from \cite{KN}, but we leave this to future work.

\bigskip

This work was supported in part by the U.S. National Science Foundation Grants No. PHY-2112729 and No. PHY-1915053 and by a PSC-CUNY grant.
\section*{Appendix A: {Basic features and geometry of ${\mathbb{CP}}^k$ spaces} }
\def\theequation{A\arabic{equation}}
\setcounter{equation}{0}

Let $t_A$ denote the generators of $SU(k+1)$ as matrices in the fundamental representation, normalized so that $\Tr (t_A t_B) = {1 \over 2} \delta_{AB}$. These generators are classified into three groups. The ones corresponding to the $SU(k)$ part of
$U(k) \subset SU(k+1)$ will be denoted by
$t_a$, $a =1, ~2, \cdots , ~ k^2 -1$ while the
generator for the $U(1)$
direction of the subgroup $U(k)$ will be denoted by 
$t_{k^2+2k}$. The $2k$ remaining generators of $SU(k+1)$ which are not in $U(k)$ are the coset generators, denoted by $t_\alpha$, $\alpha = k^2,\cdots, k^2+2k-1$. 
(To distinguish the various components, we use Greek letters from the beginning of the alphabet here; the corresponding $E$'s defined below
will be the components in the tangent frame, not the coordinate frame.
Lower case Roman letters from the middle of the alphabet onwards
will denote components in the coordinate frame, for spatial directions
only. When spacetime coordinate frames are involved, we use
Greek letters from later in the alphabet for
the coordinate frame.)


We can now use a $(k+1) \times (k+1)$ matrix $g$ in the fundamental representation of $SU(k+1)$ to coordinatize ${\mathbb{CP}}^k$, 
with the identification $g \sim gh$, where $h \in U(k)$. 
We can expand $g^{-1} dg$ which is an element of the Lie algebra as
\beqar
g^{-1}dg &=& \big( -i E^{k^2+2k} t_{k^2+2k} -i E^{a} t_{a} -i E^{\alpha} t_{\alpha} \big) \nonumber\\
&=&\big( -i E^{k^2+2k} t_{k^2+2k} -i E^{a} t_{a} -i E^{+I} t_{+I} -i E^{-I}t_{-I}\big)\label{1A}
\eeqar
where 
\beq
E^{+I} = E^{k^2+2I-2}-i E^{k^2+2I-1}, \hskip .2in
E^{-I} = E^{k^2+2I-2}+i E^{k^2+2I-1}, \hskip .2in
I=1,\cdots,k
\label{1a}
\eeq
$E^{\alpha}$ are 1-forms corresponding to the
frame fields in terms of which the Cartan-Killing metric on ${\mathbb{CP}}^k$ is given by
\beq
ds^2 = g_{ij} dx^i dx^j = E^\alpha_i E^\alpha_j dx^i dx^j
\label{2A}
\eeq
The K\"ahler one-form on ${\mathbb{CP}}^k$ is given by
\beq
\alpha = i \sqrt{2 k \over k+1} \, \Tr \left( t_{k^2 + 2k}  g^{-1} dg \right)
= \sqrt{k \over 2 (k+1)} \, E^{k^2 +2 k}
\label{2Ab}
\eeq
The corresponding K\"ahler two-form $\Omega_{\rm K} = d \alpha$ is
given by
\beqar
\Omega_{\rm K} & = & -i \sqrt{{2k \over {k+1}}} \tr \left( t_{k^2 + 2k} ~ g^{-1}dg \wedge g^{-1}dg \right)
\nonumber
\\ & = & -{1 \over 4} \sqrt{{2k} \over {k+1}} f_{(k^2+2k)\alpha\beta} ~E^\alpha \wedge E^{\beta} ~=  -{ 1 \over 4} \epsilon_{\alpha\beta} ~E^\alpha \wedge E^{\beta}
\label{3A}
\eeqar
$f_{ABC}$ are the $SU(k+1)$ structure constants, defined by $[t_A,~t_B] = i f_{ABC} \,t_C$. In deriving the last line of (\ref{3A}) we used the fact that $f_{(k^2+2k)\alpha\beta}=  \sqrt{{k+1} \over {2k}}~ \epsilon_{\alpha\beta}$, where 
\beq
\begin{array} {r c l}
\epsilon_{\alpha\beta} = -\e_{\beta \alpha} = 1&~ \hskip.2in
{\text{for}}&\alpha= k^2+2I -2,\\
 &&\beta= k^2+2I -1, \hskip .2in I=1, 2,\cdots, k\\
=  0 &&{\text{for all other choices}}
 \end{array}
 \label{3Ab}
\eeq
The volume of ${\mathbb{CP}}^k$ is normalized so that
\beq
\int_{\mathbb{CP}^k} ~\Bigl({ \Omega_{\rm K} \over {2 \pi}} \Bigr)^k = 1
\label{6A}
\eeq
The Maurer-Cartan identity $d (g^{-1} dg ) = - g^{-1} dg g^{-1} dg$,
along with (\ref{1A}), leads to
\beqar
dE^{k^2+2k} & = & -\half f^{(k^2+2k)\alpha \beta} E^{\alpha} \wedge E^{\beta} = 2 \sqrt{{k+1} \over {2k}} ~\Omega_{\rm K} \nonumber \\
dE^a + \half f^{abc} E^b \wedge E^c & = & - \half f^{a \alpha \beta} E^{\alpha}\wedge E^{\beta}
\label{7A}\\
dE^{\alpha} & = & - f^{\alpha A  \beta} E^{A} \wedge E^{ \beta}
\nonumber
\eeqar

The $U(k)$ spin connection $\omega^{IJ}$ is defined in terms of the 
holomorphic frame fields $E^{+I}$ by
\beq
dE^{+I} + { \omega}^{IJ} E^{+J} =0 ~~,~~ I=1, \cdots, k
\label{7Ab}
\eeq
$\omega^{IJ}$ takes values in the Lie algebra of $U(k)$, so one can write
\beq
\omega^{IJ} = -i \left( \omega^0 {\bf 1}  + \omega^a t_a \right)
\label{7Ac}
\eeq
The curvature two-form is given by
\beqar
R  &=&  d\omega + \omega \wedge \omega 
 =  -i \Bigl( R^0 \bold{1} + R^a t_a \Bigr)
\label{9A}\\
R^0 &=& d\omega^0   \nonumber\\
R^a &= &d\omega^a +\half f^{abc} \omega^b \omega^c \label{9a}
\eeqar
Even though equations up to (\ref{7A}) used specific properties of
$\mathbb{CP}^k$, equations (\ref{7Ab}) to (\ref{9a}) hold for any manifold with a complex structure so that the holonomy group is $U(k)$.
In deriving the effective actions including gauge and gravitational fluctuations for higher dimensional QHE we have used the topological property of the Dolbeault index to move away from the specific gauge and curvature background values. So the equations which hold in general
are (\ref{7Ab}) to
(\ref{9a}).

If we now specialize to the case of $\mathbb{CP}^k$, using Maurer-Cartan identities (\ref{7A}) we can identify the 
$\omega^{IJ}$ as
\beq
\omega = {\bar \omega}  =  -i \Bigl( \sqrt{{k+1} \over {2k}} E^{k^2+2k} ~\bold{1} +  E^a ~t_a \Bigr) \equiv  -i \Bigl( \omega^0 ~\bold{1} +  \omega^a ~t_a \Bigr)
\label{8A}
\eeq
The curvature components for $\mathbb{CP}^k$ are then given by
\beqar
\bar{R}^0 &=&  {{k+1} \over k} \Omega_{\rm K}   \nonumber\\
\bar{R}^a &= & -\half f^{a \alpha \beta} {E}^{\alpha} \wedge {E}^{\beta}
\label{background}
\eeqar
where we have indicated the background values with an overbar,
as in $\bar{R}$. Notice that in the tangent frame, the
curvatures are given in terms of the $U(k)$ structure constants. 

We can now use the freedom of $h$ transformations to parametrize $g$ in terms of complex coordinates $z^i,~\bar{z}^i$. We choose a parametrization such that 
\beq
g_{i, k+1}= {z_i \over \sqrt{1+ z \cdot \bar{z}}} ~, i=1,\cdots, k, \hskip .3in g_{k+1, k+1}= {1 \over \sqrt{1+ z \cdot \bar{z}}}
\label{5A}
\eeq
Using this parametrization one can write the K\"ahler two-form $\Omega_{\rm K}$ in (\ref{3A}) in terms of the local complex coordinates in the more familiar form
\beq
\Omega_{\rm K} = i \Bigr[ {{dz \cdot d\bar{z}} \over {1 + z\cdot \bar{z}}} -  {{\bar{z} \cdot dz ~z \cdot d\bar{z}} \over {(1 + z\cdot \bar{z})^2}}\Bigl]
\label{4A}
\eeq
We further choose the relation between the complex coordinates and the real ones to be the usual one
\beq
z^i = x^{2i-1} + i x^{2i}~~~,~~~\bar{z}^i = x^{2i-1} - i x^{2i}~~~,~i=1,\cdots,k
\label{0A}
\eeq
The parametrizations (\ref{0A}) and (\ref{5A}) determine the appropriate choice for the Cartesian frame fields in the following way.  For simplicity we will work with $\mathbb{CP}^1=S^2$ but the argument works in general. Using (\ref{5A}) one finds that the complex frame field $E^{+} = E^1-iE^2=  i dz/{(1+z\bar{z})}$. In the flat limit, where the radius of the sphere becomes large, $E^{+} = E^1-iE^2=i(e^1+ i e^2) \sim i(dx + i dy)$, where $(e^1, e^2) = (-E^2, -E^1)$. It is the $e$'s that provide the conventional
Cartesian frame fields given the choices (\ref{0A}) and (\ref{5A}). 
More generally, for $\mathbb{CP}^k$,
\beq
(e^{2 I -1} ~,~ e^{2 I }) = (-E^{k^2+2I-1}~,~ -E^{k^2+2I-2}), \hskip .3in I=1,\cdots,k
\label{4a}
\eeq
In terms of the Cartesian frame fields $e$, the
K\"ahler two-form $\Omega_{\rm K}$ and the metric $g_{ij}$ can be 
written as
\beq
\Omega_{\rm K} = { 1 \over 4} \epsilon^{\alpha\beta} ~e^\alpha \wedge e^{\beta} ,
\hskip .3in
g_{ij}  =  e^\alpha_i  e^{\alpha}_{j}
\label{4A}
\eeq
where $\alpha=1,\cdots, 2k$ and $\epsilon^{12}=\epsilon^{34}=\cdots=\epsilon^{2I-1,2I} =1$ for $I=1,\cdots,k$.


The spin connection and the corresponding curvature defined
in (\ref{7Ab}) to (\ref{9a}) involve the $U(k)$ subalgebra of the
vector representation of the Lie algebra of
the $SO(2k )$ holonomy group. 
In the effective actions for quantum Hall effect in higher dimensions we have considered fluctuations of the holomorphic $U(k)$ spin connection away from their background values. Further, in obtaining the energy-momentum tensor from the effective actions we must consider arbitrary variations
of the metric. For this, we will need to consider
the connection and curvature in $SO(2k)$. Thus 
 it is important to know how the $U(k)$ spin connection and curvature are embedded in $SO(2k)$. We will now derive this  relation for the four dimensional quantum Hall effect on $\mathbb{CP}^2$, although similar expressions hold for all $\mathbb{CP}^k$.

The real components of the $SO(4)$ spin connection can be identified via
\beq
de^{\alpha} + \omega^{\a\b} e^{\b}=0
\label{connection}
\eeq
Using (\ref{1a}),(\ref{7Ab}) and (\ref{4a}), we find
\beqar
\omega^{\a\b} &=& \epsilon^{\a\b} \omega^0~+~(J^a)^{\a\b} \omega^a \nonumber \\
R^{\a\b} &=& \epsilon^{\a\b} R^0~+~(J^a)^{\a\b} R^a 
\label{relation}
\eeqar
where $\omega^0~, R^0$ and $\omega^a~, R^a$ are the $U(1)$ and $SU(2)$ components of the complex spin connection and curvature as defined in (\ref{7Ac}), (\ref{9A}). 
The $(4 \times 4)$-matrices $J^a$ are related to the $SU(3)$ structure constants $f^{a\alpha\beta}$ via
\beq
(J^a)^{\alpha\beta} = (f^1,~-f^2,~f^3)^{3+\alpha, 3+\beta}
\label{10b}
\eeq
In particular
\beq
J^1 = \half \left[ \begin{matrix}
                               0&0&0&1\\
                               0&0&-1&0\\
                               0&1&0&0\\
                               -1&0&0&0\\
                               \end{matrix}\right], \hskip .1in
                               J^2 = \half \left[ \begin{matrix}
                               0&0&-1&0\\
                               0&0&0&-1\\
                               1&0&0&0\\
                               0&1&0&0\\
                               \end{matrix}\right] , \hskip .1in
                                J^3 = \half \left[ \begin{matrix}
                               0&1&0&0\\
                               -1&0&0&0\\
                               0&0&0&-1\\
                               0&0&1&0\\
                               \end{matrix}\right]                             
 \label{matrices}                                                             
\eeq                               
Similar expressions hold for all $\mathbb{CP}^k$ manifolds. 

The matrices $J^a$ form a basis for the Lie algebra of
$SU(2)$, obeying the commutation rules
\beq
[ J^a~,~J^b] = \epsilon^{abc} J^c 
\label{10c}
\eeq
Further, they satisfy the relations
\beqar
\epsilon^{\a\b} (J^{a})^{\b\a}&=&0~, \hskip .3in \Tr\left(J^a J^b\right) = - \delta^{ab} \nonumber \\
(J^a)^{\a\b} (J^{a})^{\gamma\delta} &=& \quarter \left(\delta^{\a\gamma}\delta^{\b\delta} -\delta^{\a\delta} \delta^{\b\gamma}\right) - {1 \over 4} \epsilon^{\a\b\gamma\delta}
\label{tildef}
\eeqar
Using (\ref{tildef}) we can write the relation between the complex $U(k)$ components of the spin connection and curvature in terms of the real $SO(2k)$ components of the corresponding quantities. In particular we have the following relations.

\noindent\underline{$\mathbb{CP}^1$ case}:
\beq
\omega^0 = \half \epsilon^{\a\b}  {\omega}^{\a\b} = {\omega}^{12}
\label{2d}
\eeq

\noindent\underline{$\mathbb{CP}^2$ case}:
\beqar
R^0 &=& \quarter \epsilon^{\a\b} {R}^{\a\b} \nonumber \\
R^a R^a &=&- 4 R^0R^0 - {R}^{\a\b} {R}^{\b\a}
\label{20A}
\eeqar 

In formulating QHE on ${\mathbb{CP}}^k$, one has to choose
the background values for the gauge fields as well.
We take the $U(1)$ and $SU(k)$ background gauge fields as
proportional to $E^{k^2+2k}_i$ and $E^a_i$. Specifically,
\beqar
\bar{A}^{k^2+2k} & = &- i n \sqrt{{{2k} \over {k+1}}} \tr (t_{k^2+2k} g^{-1} dg ) = {n \over 2} \sqrt{{{2k} \over {k+1}}} E^{k^2+2k} \nonumber \\
\bar{A}^a = E^a & = & 2i \Tr (t^a g^{-1} dg) 
\label {Abar}
\eeqar
The corresponding $U(1)$ and $SU(k)$ background field strengths are
\beq
\bar{F}  =  n\, \Omega_{\rm K} , \hskip .3in
\bar{F}^{a}  =   \bar{R}^a
\label{barF}
\eeq
We see from (\ref{barF}) that the background field strengths are proportional to the background curvature components which are constant in the appropriate frame basis, proportional to the $U(k)$ structure constants (\ref{background}). It is in this sense that the field strengths in (\ref{barF}) correspond to uniform magnetic fields appropriate in defining QHE. 

In terms of the frame fields $e^\alpha$, the curvatures for $\mathbb{CP}^2$ are given by
\begin{align}
R^0=& {3\over 2} \Omega_{\rm K} = {3\over 4} \left( e^1 e^2 + e^3 e^4\right)
\nonumber\\
R^1=& {1\over 2} \left( e^1 e^4 - e^2 e^3\right), \hskip .2in
R^2= -{1\over 2} \left( e^2 e^4 + e^1 e^3\right)
\label{CP^2curv}\\
R^3=& {1\over 2} \left( e^1 e^2 - e^3 e^4\right)
\nonumber
\end{align}
In particular, we have the relation $R^a \wedge \Omega_{\rm K} = 0$.

A few other relations, which might be of interest for
$\mathbb{CP}^k$, for arbitrary $k$, are the following. 
\beq
\int_{\mathbb{CP}^k} {\rm td}\, (T_c K) \big\arrowvert_{2k} = 1
\label{10A}
\eeq
where ${\rm td}\,(T_c K)$ is the Todd class in the complex tangent space and in 
(\ref{10A}) the $2k$-form is selected as the integrand. Explicitly,
the Todd class has the expansion given in (\ref{band14}) as
\beq
{\rm td} = 1 + {1\over 2} \, c_1 +{1\over 12} ( c_1^2 + c_2) + {1\over 24} c_1\, c_2
+ {1\over 720} ( - c_4 + c_1 c_3 + 3 \, c_2^2 + 4\, c_1^2 \,c_2 - c_1^4) + \cdots
\label{11A}
\eeq
where $c_i$ are the Chern classes. The first few Chern classes can be easily evaluated using (\ref{background}) as
\beqar
c_1 & = & \Tr ~{iR \over 2\pi} = (k+1) { \Omega_{\rm K} \over 2\pi} \nonumber \\
c_2 & = & {1\over 2} \Bigl[ (\Tr {iR \over 2\pi})^2 - \Tr ({iR \over 2\pi})^2 \Bigr] = {\half} k (k+1) ({ \Omega_{\rm K} \over 2\pi})^2 
\label{12A}
\eeqar
In deriving the expression for $c_2$ we used the fact that
\beqar
R^a \wedge R^a & = &\quarter f^{a \alpha\beta}f^{a \gamma\delta} E^{\alpha}E^{\beta}E^{\gamma}E^{\delta} = -2~{k+1 \over k}  \Omega_{\rm K}^2 \nonumber \\
\Tr \bigl[{{iR\wedge iR}} \bigr]& = & k (R^0)^2 + \half (R^a)^2 = (k+1) \Omega_{\rm K}^2
\label{13A}
\eeqar
These can be easily shown using completeness relations for the 
matrices $t^A$ in the fundamental representation.
More generally the Chern classes for ${\mathbb{CP}}^k$ can be written as
\beq
c_i =  {{k!} \over {i! (k-i)!}} \left( {\Omega_{\rm K} \over 2\pi}\right) ^i
\label{13AA}
\eeq
Using (\ref{6A}) and (\ref{12A}), we can easily check the validity of (\ref{10A}) for $\mathbb{CP}^1$, 
$\mathbb{CP}^2$ and $\mathbb{CP}^3$, the needed integrals being
\beqar
\int_{\mathbb{CP}^1} c_1 & = & 2 \int { \Omega_{\rm K} \over 2\pi} =2 \nonumber \\
\int_{\mathbb{CP}^2} c_1^2 + c_2 & = & (3^2+3) \int \left({ \Omega_{\rm K} \over 2\pi}\right)^2 =12 \nonumber \\
\int_{\mathbb{CP}^3} c_1 c_2 & = & 4 \times 6 \int \left({ \Omega_{\rm K} \over 2\pi}\right)^3 =24 
\label{14A}
\eeqar
\section*{Appendix B: An identity on determinant of $\Omega$}
\def\theequation{B\arabic{equation}}
\setcounter{equation}{0}
In this Appendix, we give a derivation of the identity (\ref{sub3}) 
for $k = 3$
and the more general case used in text.
(In what follows, $\Omega$ is the Berry curvature given in
(\ref{band4}).) consider Grassmann variables $Q_a$ and $\eta_a$,
$a= 1, 2, \cdots, 2k$, and
start with the identity
\beqar
\int [dQ] e^{Q \Omega Q } e^{\eta\cdot Q} &=& {\cal K}\,  e^{\eta \Omega^{-1} \eta /4}
\label{AppB1}\\
{\cal K} &=& \left[
{1\over k!} \e_{a_1 a_2 \cdots a_{2k} } \Omega^{a_1a_2} \Omega^{a_3 a_4}
\cdots \Omega^{a_{2k-1} a_{2k}}\right]
\nonumber
\eeqar
We equate the term with $2 l$ powers of $\eta$ on both sides.
We also carry out the integration on the $Q$'s on the left hand side
by expanding $e^{Q \Omega Q }$.
This gives us the relation
\begin{align}
\left[ {(-1)^l \over (2l)! (k-l)!}\right]&
\eta_{a_1} \cdots \eta_{a_{2l}}\,
\e_{a_1 \cdots a_{2l} a_{2l+1} \cdots a_{2k}}
\left( \Omega^{a_{2l+1} a_{2l+2}} \cdots \Omega^{a_{2k-1} a_{2k}}
\right)\nonumber\\
&= {\cal K} {1 \over 4^l l!} \eta_{a_1} \cdots \eta_{a_{2l}}\,
\left( ( \Omega^{-1})^{a_1 a_2}\cdots ( \Omega^{-1})^{a_{2l-1} a_{2l}}
\right)
\label{AppB2}
\end{align}
We can reove the $\eta$'s by writing $\left( ( \Omega^{-1})^{a_1 a_2}\cdots ( \Omega^{-1})^{a_{2l-1} a_{2l}}\right)$ in the fully antisymmetrized form,
so that
\begin{align}
\left[ {(-1)^l \over (2l)! (k-l)!}\right]&
\e_{a_1 \cdots a_{2l} a_{2l+1} \cdots a_{2k}}
\left( \Omega^{a_{2l+1} a_{2l+2}} \cdots \Omega^{a_{2k-1} a_{2k}}
\right)\nonumber\\
&= {\cal K} {1 \over 4^l l!}
\left( ( \Omega^{-1})^{a_1 a_2}\cdots ( \Omega^{-1})^{a_{2l-1} a_{2l}}
\right)_{\rm antisym}
\label{AppB3}
\end{align}
This is the basic identity. In calculating the Hall currents, we get
this expression multiplied by
a factor of $\e^{ijk \cdots a_1 \cdots a_{2l}}$. This allows us to
write the identity
\begin{align}
\left[ {(-1)^l \over (2l)! (k-l)!}\right]&
(2 l)! \,\delta^{ij \cdots }_{a_{2l+1} \cdots a_{2k}}
\left( \Omega\, \Omega \cdots\right)^{a_{2l+1} \cdots a_{2k}}\nonumber\\
&= {\cal K} {1\over 4^l l!} \e^{ij\cdots a_1  \cdots a_{2l}}
\left( (\Omega^{-1})^{a_1 a_2} \cdots (\Omega^{-1})^{a_{2l-1} a_{2l}}
\right)
\label{AppB4}
\end{align}
Because of the antisymmetry of $\delta^{ij \cdots }_{a_{2l+1} \cdots a_{2k}}$
we get all permutations of all indices in 
$\left( \Omega\, \Omega \cdots\right)^{a_{2l+1} \cdots a_{2k}}$
on the left hand side of this equation.
Since permutations of the $\Omega$'s themselves
($(k-l)!$ of these) and the permutation of the two indices on
each $\Omega$ ($2^{k-l}$ of these) do not change the expression, we can
write
\beq
\delta^{ij \cdots }_{a_{2l+1} \cdots a_{2k}}
\left( \Omega\, \Omega \cdots\right)^{a_{2l+1} \cdots a_{2k}}
= (k-l)! \, 2^{k-l} \sum_{\rm dist. perm.}
\left( \Omega\, \Omega \cdots\right)^{ ij  \cdots}
\label{AppB5}
\eeq
The number of terms in the sum in this equation is given by
\beq
{\text{Number of distinct permutations}}
= { (2k -2l)! \over 2^{k-l} \, (k-l)!}
\label{AppB6}
\eeq
Since
\beq
{1\over k!} \left( {\Omega \over 2 \pi}\right)^k 
= {1\over 2^k (2\pi)^k } {\cal K} \, d^{2k}p ,
\label{AppB7}
\eeq
we can use (\ref{AppB5}) to bring (\ref{AppB4}) to the form
\beq
\int{1\over k!} \left( {\Omega \over 2 \pi}\right)^k \,
\e^{ij\cdots a_1  \cdots a_{2l}}\left[ {(-1)^l\over l!}
\left[ \left({(\Omega^{-1})^{a_1 a_2}\over 2 (2\pi )}\right) \cdots \left({(\Omega^{-1})^{a_{2l-1} a_{2l}} \over 2 (2\pi)}\right)
\right] \right]
=\nu_{k-l}^{i j \cdots}
\label{AppB8}
\eeq
where we define
\beq
\nu_{k-l}^{ij  \cdots} = \int {1\over (2\pi)^{k+l} } \sum_{\rm dist. perm.}
\left( \Omega\, \Omega \cdots\right)^{ ij  \cdots}
\label{B9}
\eeq
Notice that there are $2k -2 l$ indices in this expression, so we can make this explicit by writing it out as
\beq
\nu_{k-l}^{i_1 i_2 \cdots i_{(2k-2l)}} = \int {1\over (2\pi)^{k+l} } \sum_{\rm dist. perm.}
\left( \Omega^{i_1i_2}\, \Omega^{i_2i_4} \cdots \Omega^{i_{(2k-2l-1)} i_{(2k - 2l )}}\right)
\label{B10}
\eeq


\end{document}